# Influence of Light on Chemical Reactivity of Water

Igor V. Shevchenko

Institute of Bioorganic Chemistry and Petrochemistry, Kiev, Ukraine. (ishev@bpci.kiev.com)

**Visible light, ultraviolet and x-ray radiation have been found to increase chemical reactivity of water. The irradiated solution of water in acetonitrile reacts with triethyl phosphite considerably faster than the non-irradiated control solution. This phenomenon is accounted for by the decomposition of water clusters under the influence of light with the formation of chemically more active free water molecules.**

It has been recently established, that the rate of interaction of triethyl phosphite with water (Fig. 1) *ceteris paribus* depends on the intensity of the Sun irradiation and on the position of the Earth with respect to the Sun and therefore displays pronounced fluctuations throughout the year. (*1*) This phenomenon was accounted for by fluctuations of the geoelectric field intensity due to the changes of impact angle of the solar radiation with respect to the Earth's atmosphere and magnetic field lines. Beside this, due to the Extreme Ultraviolet (EUV) emission of the Sun is very unstable the rate of this reaction displays also changes over short timescales.

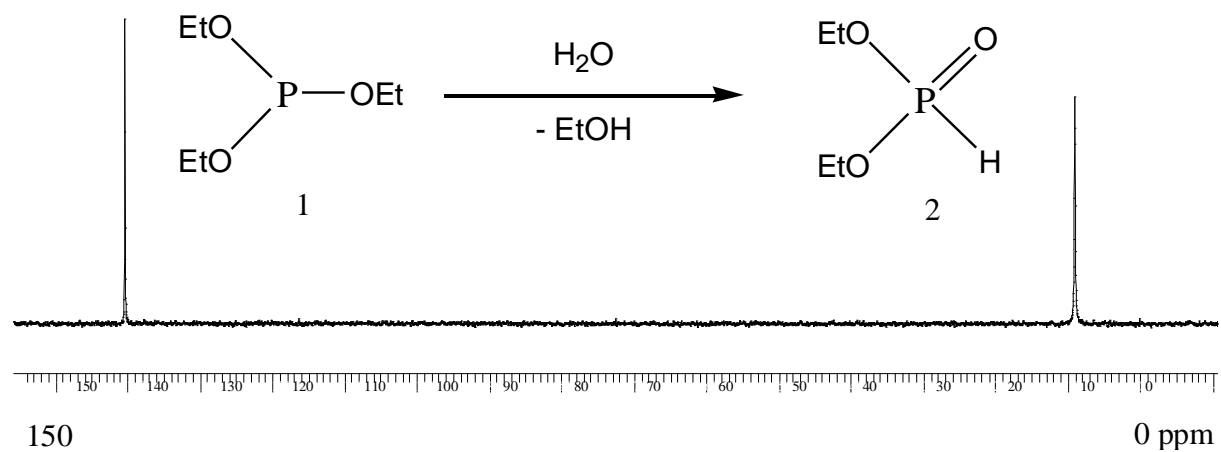

Figure 1.   Hydrolysis of triethyl phosphite **1** into diethyl phosphite **2**. $^{31}$P-NMR spectrum displays two signals at +140 ppm and +9 ppm respectively. Measuring the integral intensities of these signals allows determining the conversion rate.

Further investigations of this remarkable phenomenon were carried out in the open air, as the ferro-concrete construction of the scientific research Institute may exert a shielding influence on the intensity of geoelectric field. During these studies another fundamental phenomenon was found. It turned out that in the daytime and at night the rates of hydrolysis of triethyl phosphite strongly differ from each other. The reaction mixture prepared after sunset reacted 10-15 times slower than the same mixture made in daylight. However, it was even more surprising that the reaction prepared in daylight did not stop after sunset, but continued to proceed at the rate which had been reached before it got dark. In contrast, the reaction rate of the mixture prepared at night did not remain low after sunrise, but accelerated rapidly.

Detailed investigation of this phenomenon showed that light exerts a strong influence on the rate of this reaction. Not only daylight in the open air, but even defused indoor light noticeably accelerates this reaction compared with a control sample in darkness. Even in a common reaction solution placed in a glass tube part of which is protected from light, the reaction rate will be higher in the illuminated part.

In 1956 E. Havinga et al. described an example of photochemical acceleration of the hydrolysis of nitrophenyl phosphates and sulphates. (*2*) The authors supposed that the mechanism of the reaction may be accounted for by the excitement of nitrophenyl groups by light, which facilitates their hydrolytic cleavage from the phosphorus atom.

The results of our investigations indicate that light can affect water itself. This means that light can accelerate all chemical reactions in which water participates as a reagent. This conclusion follows from the fact that light increases the activity of water, even in the absence of a substance, reacting with it. For example, if a freshly prepared 1.7% solution of water in acetonitrile is left for 3 days in diffused daylight even in cloudy weather, after addition of triethyl phosphite the rate of hydrolysis will be 5-6 times higher (at 20 ° C) than in the same solution which was kept in darkness. The degree of activation of water depends on the radiation dose, that is, on the duration of exposure, intensity and frequency of light. Ultraviolet irradiation exerts significantly stronger influence than visible light. If, before adding triethyl phosphite, aqueous acetonitrile is irradiated with ultraviolet light (at a distance of 5 cm from 8W UV-lamp), its activity increases 5-6 times already in 2 hours, and after 24 hours of irradiation the rate of the reaction with triethyl phosphite increases 15-20 times. It is important to note, that further increase of the irradiation time does not lead to growth of water reactivity any more. (Fig. 2)

A similar strong acceleration of the reaction also occurs when the solution of water in acetonitrile before adding triethyl phosphite is exposed to X-ray irradiation.

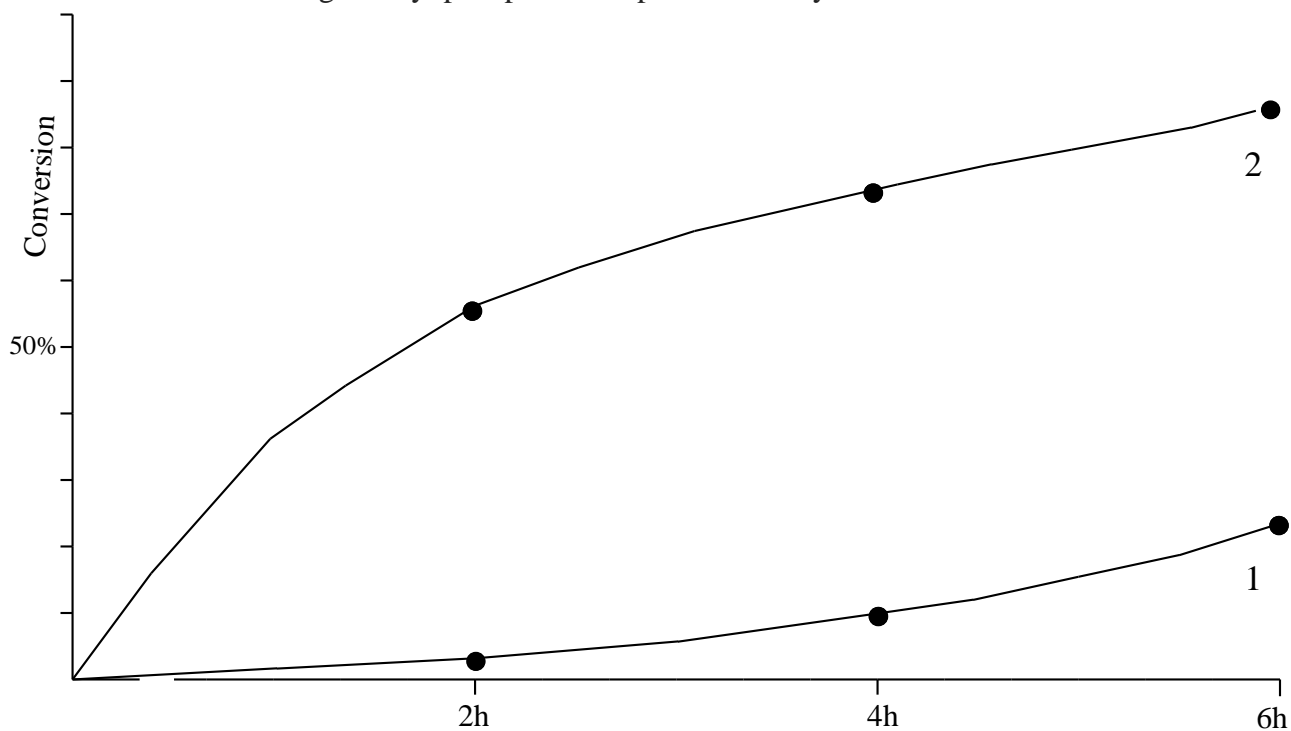

Fig.2. Rate of reaction of (EtO)3P with 1.7% solution of H2O in MeCN which was not irradiated (Line 1) and which was preirradiated for 24h at a distance of 5 cm from 8W UV-lamp (Line 2).

The reverse process of deactivation of water takes place too. After a few days in darkness the activity of the solution of water in acetonitrile irradiated by light is reduced to the original level.

An increase of chemical reactivity of water under the action of light is well accounted for by the ability of water molecules to form clusters $(H_2O)_n$, which are stabilized by intermolecular hydrogen bonds. The size of clusters may vary in a wide range – from a few water molecules to several hundreds. Water clusters were intensively studied by theoretical calculations and spectroscopic methods. (*3-14*) UV-Photoionization of neutral molecular beams of water clusters in vacuum has been also detected spectroscopically. (*15,16*)

Theoretical calculations and spectroscopic methods have shown that the distance between neighboring oxygen atoms is not constant and depends on the cluster size. This distance is longer in small clusters and shorter in large ones. (*10,17,18*) In addition, larger clusters contain fewer free hydroxyl groups –OH which are not involved in the network of hydrogen bonds. All this reduces the chemical reactivity of large clusters, whereas clusters of the size $n < 6$ are especially active, since they are unable to form three-dimensional structures. (*10,11,19-21*)

In the bulk phase all water molecules form a common continuous three-dimensional network of hydrogen bonds in which every molecule has tetrahedral bonding directions. (*10,11*). However, on mixing water with other solvents, this network disintegrates with the formation of clusters. So far this phenomenon has been known for acetonitrile and tetrachloromethane. (*12, 22*) Acetonitrile molecules are able to coordinate around water clusters. (*12*)

In our investigation a 1.7% solution of water in acetonitrile was used. After mixing bulk water with acetonitrile free water molecules and small active clusters are absent in the solution and only large three-dimensional clusters interact with triethyl phosphite after its addition. The kinetics of this reaction proves this.

In darkness the hydrolysis of triethyl phosphite in a freshly prepared solution of water in acetonitrile at first proceeds very slowly, however, over time, the rate gradually increases. This can be rationalized by the increase of the reactivity of clusters owing to the reduction of their size during the interaction with $(EtO)_3P$.

After adding triethyl phosphite to the irradiated 1.7%-solution of water in acetonitrile the kinetics of the reaction displays the opposite character. In the beginning the reaction proceeds at a high rate, which after a while decreases. It points to the fact that under the influence of light, the decrease in size of clusters occurs owing to the splitting off of single water molecules, whose high chemical activity determines the initial reaction rate.

The degree of depletion of clusters is determined by the amount of energy absorbed, that is, by the intensity and/or duration of irradiation. After the cessation of the action of light, the molecules of water which have split off combine in clusters again and the chemical activity of water decreases. However, this recombination process is slow, especially if the clusters are surrounded by a large number of acetonitrile molecules. (*23*) Therefore, a solution of water in acetonitrile, which was excited by light, retains an increased activity for a long time. This explains the observed phenomenon when the reaction mixture prepared in daylight maintained a high rate after sunset, while the reaction mixture prepared after it got dark was inert.

That it is water clusters in acetonitrile that are sensitive to light follows from the fact that bulk water itself does not display such sensitivity. Dissolution in acetonitrile of irradiated and not irradiated water gives solutions of equal activity. Bulk water does not contain clusters, but consists of molecules involved in common network of hydrogen bonds.

A detailed study of the photosensitivity of water clusters revealed another interesting phenomenon. It turned out that after the addition of triethyl phosphite, the sensitivity of water in acetonitrile to light increases approximately 100-fold. As noted above, if ultraviolet irradiation of water in acetonitrile within 2 hours increased its reactivity to triethylphosphite by 5-6 times, after addition of triethyl phospite only 1 minute of irradiation was sufficient for such acceleration. Figure 3 shows the kinetics of hydrolysis of triethylphosphite in acetonitrile at 20 $^{o}$C without irradiation (line 1) and after irradiating the reaction solution (immediately after mixing the reagents) for 1, 5 and 15 minutes (lines 2, 3 and 4). On ending the irradiation, the reaction solutions were placed in darkness, where they continued to react at different rates. It is interesting to note that after a 15-minute irradiation the reaction acquired a maximum rate, which no longer increased, even if the irradiation continued until the end of the interaction.

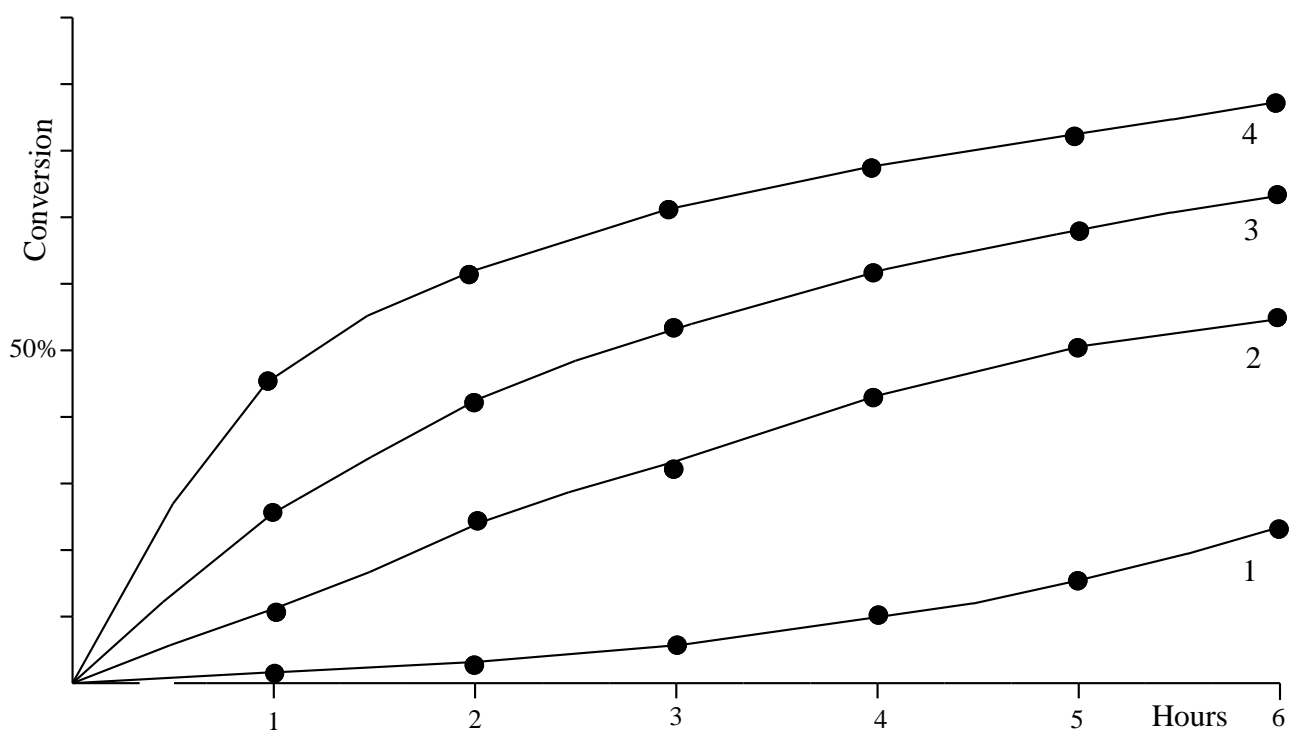

**Fig 3**. Influence of UV irradiation on the reaction mixture of (EtO)3P with 1.7% solution of H2O in MeCN. Rate of the reaction of not irradiated reaction mixture (Line 1) and initially irradiated for 1, 3 and 15 minutes (Lines 2, 3 and 4). Line 2 displays also the rate of the reaction after 5 minutes exposition of the reaction mixture to x-rays.

A sharp increase in the photosensitivity of water clusters in the presence of triethyl phosphite molecules can be explained by the formation of an intermediate complex between water cluster and triethyl phosphite molecules. The coordination of the free electron pairs of triethyl phosphite molecules with the hydroxyl groups in such a complex can probably facilitate the splitting off of water molecules from the water cluster under the action of light.

One more remarkable phenomenon consists in the fact that stability of clusters depends on the water from which they are formed. If tap water or mineral water is added to acetonitrile, the influence of light on such a solution is noticeably smaller (up to 50%) compared with an analogous solution prepared from distilled water. This is accounted for by the absence of mineral salts in distilled water. This conclusion is confirmed by the fact that dissolving 0.5% of sea salt in distilled water makes it resist light like tap water. (*24*) Beside this, the relaxation of an activated by light solution of distilled water in acetonitrile occurs considerably faster after adding small amount of sea salt.

Thus, the stability of water clusters and therefore the chemical reactivity of water is influenced by two factors simultaneously – electric field and electromagnetic oscillations of high frequency (visible light, UV-irradiation and x-rays). The influence of these factors in natural conditions is determined by the solar radiation intensity as well as by the position of the Earth with respect to the Sun. In this connection it is important to underline that the data presented in this paper were obtained from November 2015 to February 2016 and then confirmed in the same period the next year. At this time of the year at the northern latitude $50^\circ 25'$, where these studies were carried out, the reactivity of water is minimal. In another time of the year, when geoelectric field accelerates the rate of hydrolysis, the influence of light on this reaction may be less pronounced.

As water plays a crucial role in the origin and evolution of life on Earth, it is important to note one more circumstance. The change of intensity of geoelectric field and the degree of illumination levels the influence of temperature on the rate of hydrolytic processes. For example, the rate of hydrolysis of triethyl phosphite in acetonitrile at 20 $^{\circ}$C in the sun can be higher than in darkness at 50 $^{\circ}$C.

**Acknowledgments:** The author thanks Professor V.P. Kazimirov for assistance in experiments with x-ray radiation. The data reported in this paper are tabulated in the Supplementary Materials and archived at the database arXiv.org.

Materials and Methods:

The composition of the reaction mixture was the same in all experiments: 30mg triethyl phosphite dissolved in 400mg acetonitrile containing 1.7% water. Only freshly prepared solution of acetonitrile with water was used. Triethyl phosphite was distilled before use. Acetonitrile was commercial and contained 0.01% water. Triethyl phosphite was added to aqueous acetonitrile under nitrogen in a dry box. All reactions and irradiation with UV light and x-rays were carried out in 5mm NMR tubes at room temperature.

For the UV-irradiation samples were placed at a distance of 5cm from an 8 watt UV-lamp. (*25*) Exposition of the samples to x-rays was performed on a УРС-50м (URS-50m) diffractometer using $CuK_\alpha$-radiation without monochromator. The HV generator was set to the intensity 16KV/10mA.

The $^{31}P$ NMR spectra were recorded with Varian Gemini 400 MHz and JEOL FX-90Q spectrometers. The $\delta^{31}P$ chemical shifts are referenced to 85% aqueous $H_3PO_4$.